\documentstyle[11pt,equation,
epsf]{article}
\input{psfig.sty}
\begin{document}
\newcommand{\xb}{{\bf x}}
\newcommand{\yb}{{\bf y}}\newcommand{\nh}{{\hat n}}
\newcommand{\mt}{{\tilde m}}
\newcommand{\Ft}{{\tilde F}}
\newcommand{\Jt}{{\tilde J}}
\newcommand{\Lt}{{\tilde L}}
\newcommand{\dth}{\Delta \theta}
\newcommand{\dch}{\Delta \chi}
\newcommand{\dthp}{\left( \Delta \theta \right)}
\newcommand{\tr}{\rm Tr}
\newcommand{\sx}{\sigma}
\newcommand{\sxa}{\sigma_1}
\newcommand{\sxb}{\sigma_2}
\newcommand{\pha}{\phi_1}
\newcommand{\phb}{\phi_2}
\newcommand{\phpa}{\phi^+_1}
\newcommand{\phpb}{\phi^+_2}
\newcommand{\phm}{\phi^-}
\newcommand{\php}{\phi^+}
\newcommand{\psif}{{\psi_f}}
\newcommand{\psa}{\psi_1}
\newcommand{\psb}{\psi_2}
\newcommand{\Phib}{\bar{\Phi}}
\newcommand{\mpl}{m_{Pl}}
\newcommand{\Mpl}{M_{Pl}}
\newcommand{\lx}{\lambda}
\newcommand{\Lx}{\Lambda}
\newcommand{\kx}{\kappa}
\newcommand{\ex}{\epsilon}
\newcommand{\be}{\begin{equation}}
\newcommand{\ee}{\end{equation}}
\newcommand{\een}{\end{subequations}}
\newcommand{\ben}{\begin{subequations}}
\newcommand{\beq}{\begin{eqalignno}}
\newcommand{\eeq}{\end{eqalignno}}
%% macros to produce the symbols "less than or of order of"
%% and "greater than or of order of" %
\def \lta {\mathrel{\vcenter
     {\hbox{$<$}\nointerlineskip\hbox{$\sim$}}}}
\def \gta {\mathrel{\vcenter
     {\hbox{$>$}\nointerlineskip\hbox{$\sim$}}}}
\pagestyle{empty}
\noindent
%July 1999
\begin{flushright}
%SNS--PH/1999--11
%\\
%hep--ph/9908209
\end{flushright} 
\vspace{3cm}
\begin{center}
{ \Large \bf
The World as a Dual Josephson Junction
} 
\\ \vspace{1cm}
{\large 
N. Tetradis 
} 
\\
\vspace{1cm}
{\it
Scuola Normale Superiore, Piazza dei Cavalieri 7, 
56126 Pisa, Italy 
} 
\\
\vspace{2cm}
\abstract{
We examine some of the implications of the field-theoretical mechanism
for the localization of gauge fields on hypersurfaces 
in higher-dimensional bulk space-time.
This mechanism exploits the analogy between confinement and dual
superconductivity. In the simplest case of a photon localized on 
a (2+1)-dimensional surface in
a (3+1)-dimensional bulk, we argue that the system behaves like
a dual Josephson junction. This implies that the effective gauge theory on
the surface is not free, but displays weak confinement with a linear 
potential. We comment on the relevance of our results for the realistic 
case of a (3+1)-dimensional surface in a space-time with one or more
extra dimensions.
\\
\vspace{1cm}
%PACS number: 98.80.Cq
} 
\end{center}
\vspace{4cm}
\noindent
%CERN--TH/97--XXX \\
%July 1999

\newpage

\pagestyle{plain}
\setcounter{page}{1}

\setcounter{equation}{0}

The possibility that the particles of the Standard Model
are localized on a (3+1)-dimensional hypersurface in a 
higher-dimensional bulk space-time has been considered repeatedly 
in the past. The localization of fermions \cite{rubshap}
relies on an index theorem  \cite{index}
that guarantees the presence of fermionic zero modes at the 
points where a background scalar field, with a Yukawa
coupling to the fermions, vanishes.
This mechanism can be employed in order to simulate
chiral fermions on the lattice by attaching them to a (3+1)-domain wall 
in the (4+1)-dimensional bulk \cite{kaplan}. The
efficiency of the mechanism has been verified through
lattice simulations \cite{chen}.

The localization of gauge fields, which will be our subject of
interest, is more difficult to achieve in a field-theoretical 
context. The most promising proposal
\cite{barnkan}--\cite{sav1} suggests that 
the low-energy gauge fields are trapped 
at the center of a defect by being massless there, while they 
become very massive in the bulk. In order to guarantee that the
localized fields are long-ranged within the defect, the effective mass 
must be generated by embedding them in 
the gauge sector of a theory that is confining in the bulk.
This mechanism has been employed in the recent 
investigations of the possible presence of
extra dimensions experimentally accessible
at TeV energies \cite{sav1,sav2}. Other aspects have been explored
in refs. \cite{other}.

In this work we examine in detail some of the implications of 
the gauge-field localization mechanism. 
Our discussion is limited to the field-theoretical scenario\footnote{
Within string theory, matter and gauge fields can be 
localized on D-branes \cite{dbranes}, with the emergence of
low-energy (3+1)-theories, weakly coupled to the modes 
propagating in the extra dimensions \cite{locbran}.
The possibility of experimentally accessible extra dimensions 
can be considered within this framework as well \cite{savign}.
}. We consider the simplest case of a $U(1)$ gauge symmetry.
Similarly to 
refs. \cite{giashif,sav1}, we discuss explicitly the reduction of
a (3+1)-dimensional theory to a lower-dimensional one. We are forced to do
so by the nature of the reduction mechanism, which involves strongly 
coupled regimes of gauge theories. As only a limited understanding of these
regimes is available in 3+1 dimensions, 
with no concrete generalization to
higher dimensions, we limit ourselves to the most accessible case.
The implications
for the reduction of a higher-dimensional theory to a (3+1)-dimensional one
can be inferred only by analogy.

Firstly we review how the photon of the $U(1)$ gauge symmetry is localized on
the brane. We then argue that, in analogy with the behaviour in
standard Josephson junctions, the resulting (2+1)-dimensional 
theory is not free. The equation of motion for the electromagnetic field
includes a small effective mass term that indicates the presence of
a finite correlation length. Moreover, configurations 
exist that resemble lines of electric flux. The most consistent 
interpretation of this behaviour is that the system displays weak confinement
with a linear potential.

The situation of interest is schematically presented in fig.~1.
Within a (3+1)-dimensional bulk space-time we consider  
a brane\footnote{
We borrow string-inspired terminology, even though our discussion
is limited to the field-theoretical scenario.}
of finite thickness $d$ along the $z$-axis. 
We assume that the 
gauge field remains massless at tree level 
within this region, while it develops 
a large effective mass in the bulk. At energy scales below this 
mass we expect an effective (2+1)-dimensional theory to
appear on the brane. A first guess would be that this theory
is a free $U(1)$ gauge theory with a massless photon. 
However, the nature of the reduced theory
depends on the mechanism through which the gauge field develops an effective
mass in the bulk.

An implementation of the above scenario \cite{barnkan} is 
obtained in the context of the Abelian Higgs model, described by the
tree-level Lagrangian 
\be
{\cal L}_{tr} = \int d^4x~\left\{
-\frac{1}{4} F_{\mu \nu} F^{\mu \nu} 
+
\left[ \left( \partial_\mu + iqA_\mu\right) \phi\right]^*
\left[ \left( \partial^\mu + iqA^\mu\right) \phi\right]
-V(\phi) \right\},
\label{lagr} \ee
with $F_{\mu\nu}=\partial_\mu A_\nu-\partial_\nu A_\mu$. We work
in Minkowski space with metric $(+,-,-,-)$.
In a simple configuration that could localize the gauge field,
the scalar field $\phi$  has
a zero expectation value inside the
brane and a constant non-zero value in the bulk. It is not important
how such a configuration is generated dynamically. It may result from
the interaction of $\phi$ with other fields \cite{barnkan}.
For our discussion we shall assume that $\phi=0$ for $|z|< d/2$ and
$\phi=\rho \exp(i\theta)$ for $|z| \geq d/2$. As a result, the gauge 
field becomes massive in the bulk through the Higgs mechanism, while
it remains massless inside the brane.

\setlength{\unitlength}{1cm}
\begin{figure}[t]
\begin{center}\hspace{-5mm}
\begin{picture}(17,9)
\put(1.5,-4){\includegraphics{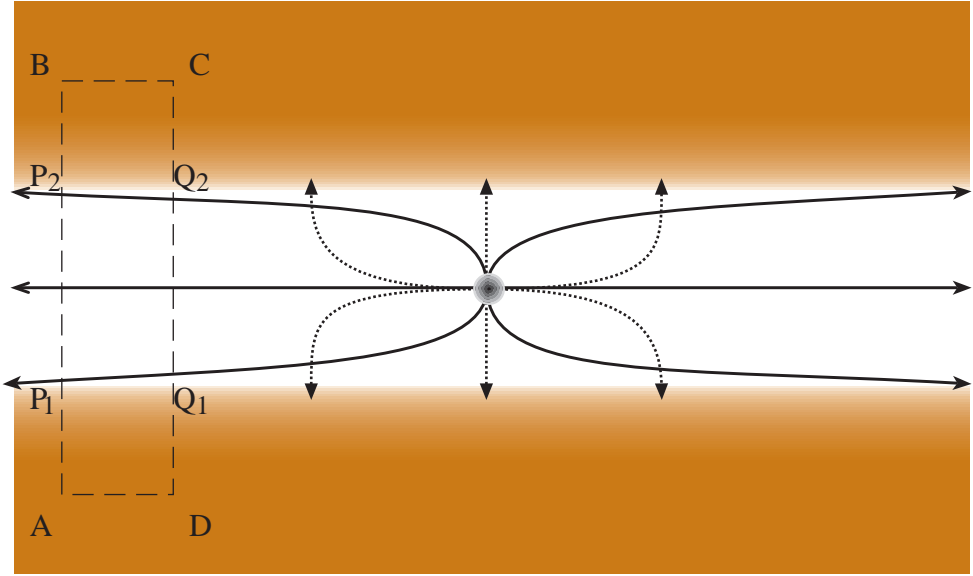}}
\end{picture}
\caption[SP]{\em 
Electric (dotted) and magnetic (solid) field lines generated by
a fictitious dyon (electrically and magnetically charged particle) 
in the Josephson junction. In the dual picture the roles of
electric and magnetic fields are reversed. The 
$x$-axis is horizontal and the $z$-axis vertical. 
\label{fig1}}
\end{center}\end{figure}

The Abelian Higgs model in the spontaneously broken phase is
equivalent to the Ginzburg-Landau theory of superconductivity. 
A well known property of superconductors is the existence of 
frictionless currents that can flow without a potential. 
For the tree-level Lagrangian of eq.~(\ref{lagr}) the current density is
given by 
\be 
J_{tr}^\mu= -2 q \rho^2 \partial^\mu \theta - 2 q^2 \rho^2 A^\mu.
\label{current} \ee
The phase $\theta$ is unobservable for a bulk superconductor,
as it can be eliminated through a gauge transformation.
The superconducting currents flow near the surface of the material
and expel magnetic fields from it. This property, the Meissner effect,
can be understood also as a consequence of the photon mass $\sqrt{2}q \rho$.
The magnetic field decays over a distance $\lx \sim (q \rho)^{-1}$. Deep
inside the superconductor, the gauge field is zero and the current
of eq.~(\ref{current}) vanishes.

The configuration of fig.~1 corresponds to a 
Josephson junction: two superconducting regions
separated by a thin layer of non-superconducting material. 
The phase $\theta$ cannot be eliminated completely in
this case. More specifically, 
we can define the gauge-independent phase difference between 
two points \cite{joseph}
\be
\dth_{P_1 P_2} 
= \theta(P_2)-\theta(P_1) - q\int_{P_1}^{P_2} \vec{A}\cdot d\vec{l}.
\label{dth} \ee
The gauge-dependent phases 
$\theta(P_1)$ and $\theta(P_2)$ can be eliminated for convenience through
an appropriate gauge transformation. The phase difference $\dth_{P_1 P_2}$
between two
points on either side of the junction can be non-zero \cite{joseph}.
Of particular interest to us will be the 
phase difference generated by the existence of 
an electromagnetic field in the non-superconducting region (see below). 

As a result of the presence of a phase difference, a superconducting 
current flows across a junction, the Josephson current.
This can be shown on general grounds \cite{weinberg} by noticing that,
beyond tree level, the effective action of the system 
depends on $\dth$ through quantum effects (tunnelling of charges
across the junction). 
The current can be obtained from the effective matter 
action through differentiation with respect to the gauge field.
The dependence of the action on a phase difference given by 
eq.~(\ref{dth}) immediately leads to the presence of a current. This 
tunnelling current does not require 
a potential difference between the two superconducting regions in order
to flow.

We can change each of the phases $\theta(P_1)$, $\theta(P_2)$ by a multiple
of $2\pi$ without altering its physical significance. This means that the
effective action and, therefore, the Josephson current must be a periodic
function of $\dth$. We parametrize the current density 
as \cite{joseph,weinberg}
\be
J^3(\dth) = J^3_{\max} \sin(\dth),
\label{jz} \ee 
where $J^3_{\max}$ is its maximum value.
We emphasize that $J^3$ is a tunnelling current. 
(The classical current of eq.~(\ref{current}) is zero in the brane.) 
This makes its calculation 
very difficult, as it depends on the form
of the barrier that separates the two bulk regions with non-zero condensates.
However, eq.~(\ref{jz}) is sufficient for our discussion. The 
maximum value $J^3_{\max}$ can be taken as a phenomenological parameter that 
can be very small 
in units of the typical mass scale of the potential
$V(\phi)$ in eq.~(\ref{lagr}). 

We can now consider the behaviour of the electromagnetic field inside
the brane. We employ the Maxwell equations 
in the presence of an external current density given by eq.~(\ref{jz}).
They correspond to the equations of motion
derived from the tree-level Lagrangian of eq.~(\ref{lagr}) with $\phi=0$ and
an external current. 
The presence of the bulk superconductors imposes certain 
conditions on the solutions of these equations. 
The electric field parallel to a conductor is zero near its surface.
As pointed out in ref.~\cite{barnkan,giashif},
this means that electric field lines
must end perpendicularly to the boundary of the Josephson junction.
For a point charge, the electric field (whose lines are 
denoted by the dotted lines in fig.~1) dies off within a distance
$ \sim d$ in the $x,y$-directions. 
As we are interested in the low-energy behaviour of the 
system, we do not consider configurations with variations of the fields
at short distances. This means that we can approximate $E_x$, $E_y$ as zero 
and assume that $E_z$ is independent of $z$ inside the brane.
The magnetic field has a continuous $z$-component at the surface of 
a conductor. As it is zero inside the superconductor, we 
assume that $B_z$ vanishes everywhere. The components $B_x$, $B_y$
are non-zero in the brane and vanish exponentially within a distance 
$\lx \sim (q \rho)^{-1}$ in the bulk. The magnetic field lines are
localized inside the brane (solid lines in fig.~1).

In terms of the gauge field $A^\mu$ the above conditions 
can be written as 
\beq 
E_x=&-\frac{\partial A^0}{\partial x}-\frac{\partial A^1}{\partial t}=0
~~~~~~~~~~
E_y=-\frac{\partial A^0}{\partial y}-\frac{\partial A^2}{\partial t}=0
~~~~~~~~~~
E_z=-\frac{\partial A^3}{\partial t}
\nonumber \\
B_x=&\phantom{-~ }\frac{\partial A^3}{\partial y}
~~~~~~~~~~~~~~~~~~~~~~~~~
B_y=-\frac{\partial A^3}{\partial x}
~~~~~~~~~~~~~~~~~~~~~~~~~
B_z=-\frac{\partial A^1}{\partial y}+\frac{\partial A^2}{\partial x}=0,
\label{amu} \eeq
where again we have assumed no $z$-dependence.
It is clear that the emerging low-energy theory does not meet our expectations.
The only unconstrained component of the gauge field is $A^3$, precisely 
the one we would like to eliminate. However, it is instructive 
for the following to take one more step and derive the equation of motion
of $A^3$. 

In order to do so, we study the behaviour of the gauge-invariant
phase difference $\dth$ along the brane.
We consider the path ABCD depicted in fig.~1. 
From eq.~(\ref{dth}) with $\theta=0$ we obtain
\beq
\dth_{AB} = &~-q\int_{A}^{B} \vec{A}\cdot d\vec{l}
~\simeq~ \dth_{P_1P_2} = -q d A^3(P)
\nonumber \\
\dth_{DC} = &~-q\int_{D}^{C} \vec{A}\cdot d\vec{l}
~\simeq~ \dth_{Q_1Q_2} = -q d A^3(Q)
\nonumber \\
\dth_{BC} = &~\dth_{DA}=0. 
\label{dthabcd} \eeq
Here $A^3(P)$, $A^3(Q)$ are the almost constant values of
the gauge field inside the brane. 
We have neglected the contribution from
the region of width 
$\lx \sim (q \rho)^{-1}$, 
in which the gauge field falls off exponentially fast 
to zero. The superconducting currents in this region and, according to
eq. (\ref{current}), 
the gauge field are parallel to the surface of the brane
to a good approximation. Along the line segments BC and DA the gauge
field is zero. From the above equations we obtain
\be
\frac{\partial \dthp}{\partial x} =
\frac{\dth_{DC}-\dth_{AB}}{\Delta x} \simeq
- q d \frac{A^3_{Q}-A^3_{P}}{\Delta x} = q d B_y.
\label{dthx} \ee
By considering a path on the $yz$-plane and repeating the above steps
we find
\be
\frac{\partial \dthp}{\partial y} = -q d B_x.
\label{dthy} \ee
Finally, taking the time derivative of the first of eqs.~(\ref{dthabcd})
we find
\be
\frac{\partial \dthp}{\partial t} = q d E_z.
\label{dtht} \ee

The above equations, eq.~(\ref{jz}) 
and the Maxwell equations give
\be
-qd \left( \frac{\partial B_y}{\partial x}-\frac{\partial B_x}{\partial y}
-\frac{\partial E_z}{\partial t}  \right) + qd J^3=
\partial^i\partial_i \dthp + q d J^3_{\max} \sin ( \dth) =0,
\label{eomdthsine} \ee
where $i=0,1,2.$ 
We conclude that there is a mode on the brane that obeys the 
sine-Gordon equation. It is obvious from eqs.~(\ref{dth}),
(\ref{dthabcd}) that this mode corresponds to the third component of
the gauge field: $\dth = -q d A^3$.
We consider weak fields $(\dth \ll 1)$, 
for which we can approximate  
eq.~(\ref{eomdthsine}) as 
\be
\left[ \partial^i\partial_i + m^2 \right] \dth =0,
\label{eomdth} \ee
with $m^2=qdJ^3_{\max}$. 
In realistic Josephson junctions, 
eq.~(\ref{eomdth}) implies the presence of a Meissner effect 
even in the non-superconducting material \cite{joseph}. 
Applied electromagnetic fields
decay over a distance 
$\sim \left(q d J^3_{\max}\right)^{-1/2}$. This phenomenon has
been observed experimentally. The decay length in the junction can
be orders of magnitude larger than the decay length in the
superconductor.
Also solitonic configurations can appear, 
corresponding to 
solutions of eq.~(\ref{eomdthsine}) \cite{joseph}.

From the point of view of obtaining an effective 
(2+1)-dimensional theory of electromagnetism, our attempt has
failed. The 0,1,2-components of the gauge field are strongly constrained
by eqs.~(\ref{amu}), while the 3-component propagates freely but has
a small mass. 
A possible remedy for this situation is provided by the suggestion
of refs.~\cite{barnkan,giashif}. 
The material in the bulk must be a {\it dual}
superconductor \cite{dualsup}. 
In other words, there must be a condensate of magnetic charge in
the bulk. 

It is believed that dual superconductivity is realized in the 
confining phase of gauge theories. The particular implementation 
of ref.~\cite{giashif} employs
an $SU(2)$ gauge theory coupled to a scalar field in the
adjoint representation (the Georgi-Glashow model) 
\cite{georgi}. Inside the brane 
the $SU(2)$ symmetry is broken down to $U(1)$ through a 
non-zero expectation value of the scalar field. The low-energy theory is in
the Coulomb phase and a  massless photon 
should emerge\footnote{In fact, the electromagnetic field of the 
(2+1)-dimensional theory develops 
a small mass due to instanton effects and electric charge becomes confined
with a linear potential
\cite{polyakov}. However, this is related to the fact that the $U(1)$
symmetry is compact in the Georgi-Glashow model and has nothing to do
with the phenomenon we discuss in this paper. Our approach is more general and
our conclusions apply to non-compact $U(1)$ as well.}.
In the bulk the scalar field has a zero expectation value and
the theory is in the confining phase. 
All excitations are very massive, and this prevents the photon that is 
localized on the brane from entering the bulk.
The connection with monopole condensation in the confining phase 
can be seen
through 't Hooft's Abelian projection \cite{thooft}.
Any operator in the adjoint representation,
such as the scalar field in this model, can be diagonalized through a
gauge transformation that preserves an ``electric'' 
$U(1)$ gauge symmetry.
This gauge transformation is singular at the points where the 
operator is zero. As a result, the zeros of various operator configurations
can be considered as the 
world lines of ``magnetic'' monopoles. The latter 
are assumed to condense and confine ``electric'' charge. 
Support for the picture of monopole condensation in the confining phase
has been obtained through lattice simulations
\cite{abelproj}.

In our discussion we shall use only the main elements
of the above picture, since its details are not well understood. 
We consider electromagnetism in the presence of $U(1)$ magnetic charge. 
We assume that a 
magnetic condensate forms in the bulk, with the appearance of
frictionless currents. In the absence of electric charge, we
can use a phenomenological local Lagrangian description
\be
{\tilde{\cal L}}_{tr} = \int d^4x~\left\{
-\frac{1}{4} \Ft_{\mu \nu} \Ft^{\mu \nu} 
+
\left[ \left( \partial_\mu + ig C_\mu\right) \psi\right]^*
\left[ \left( \partial^\mu + ig C^\mu\right) \psi\right]
-V(\psi) \right\}.
\label{lagrd} \ee
The dual gauge field $C^\mu$ is defined through the relation 
\cite{zwanziger,report}
\be
\Ft^{\mu \nu} =\frac{1}{2} \epsilon^{\mu \nu \lambda \sigma} 
F_{\lx \sigma}= \partial^\mu C^\nu-\partial^\nu C^\mu, 
\label{dualfield} \ee
and $g$ is the magnetic charge.
A frictionless magnetic current flows near the surface of the
regions of non-zero expectation value for
the magnetic condensate
 $\psi=\sigma \exp(i\chi)$.
At tree level it is given by 
\be 
\Jt_{tr}^\mu= - 2 g \sigma^2 \partial^\mu \chi - 2 g^2 \sigma^2 C^\mu.
\label{currentd} \ee
A dual Meissner effect prevents the electric field from entering 
the regions with $\psi\not= 0$. 

The system of fig.~1 can be viewed now as a dual Josephson junction 
with a tunnelling magnetic current $\Jt^3$ flowing across the brane. 
It is clear from eq.~(\ref{currentd})
that the gauge-invariant definition of the phase
difference $\dch$ between the two sides of the brane must involve the
dual field $C^\mu$. 
Repeating the arguments that led to eq.~(\ref{jz}) we find
\be
\Jt^3(\dch) = \Jt^3_{\max} \sin(\dch).
\label{jzd} \ee 
We emphasize at this point 
that the presence of a current is independent of the detailed form of the
Lagrangian of the system. It is a consequence only of our assumption
that a condensate exists on either side of the brane 
\cite{weinberg}. 
This is an important point, because a consistent Lagrangian 
description of a system with both electric and magnetic charges will be
much more complicated than eq.~(\ref{lagrd})
(and probably non-local) \cite{report}. 
However, we believe
that our arguments should be valid in the general case as well.

We turn next to the gauge field localized
on the brane. 
We discuss its behaviour starting from the 
Maxwell equations, which we assume to be the correct equations of motion.
It is sufficient for our discussion to consider a
propagating electromagnetic field in the absence of electric
charges or currents.  
The Maxwell equations read 
\beq
\partial_\mu F^{\mu\nu} =& 0,
\label{maxwell1} \\
{\partial_\mu} \Ft^{\mu\nu} =& \Jt^\nu.
\label{maxwell2} \eeq
In the second equation we have included 
the tunnelling magnetic current $\Jt^3$.
We must also take into account the constraints on the
electromagnetic field arising from the presence of the dual
superconducting phase in the bulk. The arguments we gave in the case of the
standard Josephson junction can be repeated with the exchange of
the role of electric and magnetic fields.

Let us assume for a moment that $\Jt^\nu=0$. Then eqs.~(\ref{maxwell1}), 
(\ref{maxwell2}) can be solved in terms of the 
gauge field $A^\mu$, as in our earlier discussion.
The constraints from the presence of the dual superconducting phase in the 
bulk now give
\beq 
E_x=&-\frac{\partial A^0}{\partial x}-\frac{\partial A^1}{\partial t}
~~~~~~~~~~~~~~
E_y=-\frac{\partial A^0}{\partial y}-\frac{\partial A^2}{\partial t}
~~~~~~~~~~~~~~
E_z=-\frac{\partial A^3}{\partial t}=0
\nonumber \\
B_x=&\phantom{-~ }\frac{\partial A^3}{\partial y}=0
~~~~~~~~~~~~~~~~~~
B_y=-\frac{\partial A^3}{\partial x}=0
~~~~~~~~~~~~~~~~~~
B_z=-\frac{\partial A^1}{\partial y}+\frac{\partial A^2}{\partial x},
\label{amud} \eeq
where again we have assumed no $z$-dependence.
It is clear that $A^3$ is not a dynamical degree of freedom of the
low-energy theory. The remaining components should belong to an effective
(2+1)-dimensional gauge theory. 

Taking into account the current $\Jt^\nu$ in eq.~(\ref{maxwell2})
forbids a simple solution in terms of $A^\mu$. 
One could try the ansatz
\cite{report}
\be
F^{\mu\nu}=\partial^\mu A^\nu-\partial^\nu A^\mu 
+ \epsilon^{\mu \nu \lambda \sigma} G_{\lx \sigma}.
\label{ansatz} \ee
Eq.~(\ref{maxwell2}) then gives
\be
-\partial^\mu \left( G_{\mu \nu}-G_{\nu \mu} \right) =\Jt_\nu.
\label{gmn} \ee
The diagonal components of $G^{\mu\nu}$ 
do not contribute to eq.~(\ref{ansatz}),
and we take them to be zero.
In our case the only non-zero component of the current is $\Jt^3$ and
we are looking for a solution with no $z$-dependence.
This suggests $G^{i j}=0$, with $i,j=0,1,2$. Notice that we cannot
take $G^{i j}$ to be symmetric, because it would not contribute to 
eq.~(\ref{ansatz}).
This leaves us with $G^{i3}$, $G^{3i}$ as possible non-zero components.
An immediate consequence is that the expressions for $B_x$, $B_y$, $E_z$
in eqs.~(\ref{amud}) are not modified, and $A^3$ decouples from
the low-energy theory.

The Maxwell equations can be solved easily in terms of the dual 
gauge field $C^\mu$
defined in eq.~(\ref{dualfield}). 
The non-vanishing components of the electromagnetic field are 
\be
E_x=-\frac{\partial C^3}{\partial y}
~~~~~~~~~~~~~~~
E_y=\frac{\partial C^3}{\partial x}
~~~~~~~~~~~~~~~
B_z=-\frac{\partial C^3}{\partial t},
\label{bmub} \ee
where we have assumed no $z$-dependence.
The reasoning that led to eq.~(\ref{eomdth}) now gives
\be
-gd \left( -\frac{\partial E_y}{\partial x}+\frac{\partial E_x}{\partial y}
-\frac{\partial B_z}{\partial t}  \right) + gd \Jt^3=
\partial^i\partial_i \left( \dch \right) + g d \Jt^3_{\max}
\sin \left( \dch \right) =0.
\label{eomdchsine} \ee
For $\dch \ll 1$ we obtain 
\be
\left[ \partial^i\partial_i + \mt^2 \right] \dch =0,
\label{eomdch} \ee
with $\mt^2=g d \Jt^3_{\max}$.
The massive mode $\dch$
can be identified with the third component of
the dual field: $\dch=-gdC^3$.

In summary, the following picture emerges: A (2+1)-dimensional
theory appears on the brane. Its description
in terms of the gauge field 
$A^i$, with $i=0,1,2$, is complicated. However, in terms of the dual
field $C^3$, the physical quantities 
$E_x$, $E_y$, $B_z$ are given by the simple expressions~(\ref{bmub}).
The field $C^3$ is massive, with a mass $\mt$ that can
be very small in units of the typical scale of the theory
in the bulk. As a result, the electromagnetic field $E_x$, $E_y$, $B_z$
has a finite correlation length.
This can be seen
by simply taking $y$, $x$, $t$-derivatives of
eq.~(\ref{eomdch}) and remembering that $\dch=-gdC^3$.
Indirect experimental support of this picture comes from
the observation of a Meissner effect in standard Josephson
junctions. 

We expect the above conclusions to remain valid in a theory that includes
electric charges on the brane. As we mentioned earlier, the 
Josephson effect is an immediate consequence of the 
presence of charged condensates and does
not depend on the details of the underlying theory \cite{weinberg}.
Therefore, the complications encountered in constructing a
consistent theory of electric and magnetic charges are not expected
to lead to significant modifications of our arguments. 
We mention that, in a consistent theory, electric and magnetic 
charges must satisfy Dirac's quantization condition
\cite{dirac}
\be
q g = 2\pi n.
\label{quant} \ee

An interesting solution of the sine-Gordon equation~(\ref{eomdchsine}) 
is given by \cite{joseph}
\be
\dch = 2 \sin^{-1} {\rm sech} \left[ \mt \left(x -x_0 \right)\right].
\label{fluxon} \ee
It corresponds to a defect localized near the line $x=x_0$. The electric
field $E_y$ is non-zero near $x_0$ and vanishes
at distances $\gta \mt^{-1}$ away from it.  
The phase $\dch$ changes by $2\pi$ as $x$ goes from $-\infty$ to 
$\infty$. We consider a defect at the center of the surface bounded by
the path ABCD in fig.~1. By calculating the line integral for the
dual gauge field along ABCD, it is easy to see that the defect carries
unit electric flux $2\pi/g = q$, for $n=1$ in eq.~(\ref{quant}). 
The energy per unit length of the 
defect is $\sim \left( \Jt^3_{\max}/g^3d\right)^{1/2}$ \cite{joseph,barone}.
Lines with larger electric flux correspond to solutions for which
$\dch$ varies by multiples of $2\pi$. It is conceivable that flux-carrying 
lines
of this type may connect opposite electric charges on the brane.

There is a close similarity between the bulk and the thin
region of width $d$ that we are considering. 
On the brane the electromagnetic field is massive and defects exist that
carry electric flux, similarly to the behaviour in the bulk. 
The most consistent interpretation of the emerging (2+1)-dimensional theory is 
that it displays confinement with a linear potential, but with a typical 
scale much smaller than the scale characterizing the theory in the bulk.
This is in agreement with 
the experimental studies of standard Josephson junctions.
In that case, the system behaves as if the
superconducting properties extend 
over the whole structure including the barrier \cite{barone}.
In a certain sense, this is caused by 
the electric condensate penetrating the barrier instead of ending
abruptly at the surface. 
For the dual picture that we are considering, we expect the magnetic
condensate to behave in an analogous way. The implication is 
that dual superconducting behaviour, and therefore confinement, 
must be present inside the brane as well.

The relevance of the localization mechanism we discussed
for the reduction of
a (4+1)-dimensional theory to a (3+1)-dimensional one is not 
firmly established. The problem arises because
the mechanism 
makes explicit use of four-dimensional electric-magnetic duality.
It is not obvious how this duality can be generalized in higher
dimensions. For this reason we do not attempt here a quantitative 
discussion of the realistic 
case. However, our basic argument, leading to the existence of the 
Josephson current, relies solely on the presence of a charged condensate
on both sides of the brane.
If the localization mechanism 
involves shielding of fields by 
frictionless currents in a
charged condensate, this argument is very likely to remain valid. 
It seems reasonable to speculate that 
the penetration of the barrier by the condensate will result in 
behaviour similar to that in the bulk, with observable consequences 
such as an effective mass for the gauge field or chanelling of flux lines
into tubes.

The most obvious experimental signature for the scenario we considered
would be the generated
effective mass for the electromagnetic field. The presence 
of flux lines between electric charges is probably too difficult to detect
because of the smallness of their energy per unit length.
The scale of new physics, related to 
the width $d$ of the brane and the confinement scale in the bulk, is
expected to be at least of the order of TeV \cite{sav1,sav2}. The experimental
upper bound on the photon mass is at the level of $10^{-16}$ eV (or possibly
10 orders of magnitude more stringent) \cite{massphoton}. This means
that the effective current density $\Jt^3_{\max} d$ across the brane 
must be $ \lta 10^{-56}$ in units of the scale of new physics.
As this current is a tunnelling one, such small values are not 
unnatural. The most exciting aspect of this scenario is that it has 
experimental low-energy implications for a non-gravitational sector. 
It demonstrates that the presence of extra dimensions 
can be unveiled much below the energies at which particles
are released in the bulk. 

We finish with a remark on the scenario with more than one 
extra dimensions. In the models of refs.~\cite{sav1,sav2} the gauge fields 
are localized inside a vortex instead of a brane. The mechanism is analogous
to the localization of magnetic fields in superconducting rings
or macroscopic 
holes in superconducting materials. One important property of such
geometries is that magnetic fields produced by a source do not
disappear when this source is removed. Instead, they are maintained
by frictionless currents at the surface of the superconductor. 
This is a more explicit manifestation of the presence of a
condensate in the bulk. If the analogy with superconductivity persists
for the reduction of a (5+1)-dimensional theory to a (3+1)-dimensional
one, a clear experimental signature will be the time delay in the
variation of electromagnetic fields with respect to the variation of
the source that generates them.

\vspace{0.5cm}
\noindent
{\bf Acknowledgements}: 
I would like to thank R. Barbieri, A. Di Giacomo, 
G. Dvali, E. Kiritsis, K. Konishi, B. Lucini, 
R. Rattazzi and A. Strumia for useful discussions.
This research was supported by the E.C.
under TMR contract No. ERBFMRX--CT96--0090.

%\newpage

\end{document}